\documentclass[aps,prd,reprint,nofootinbib,showkeys,showpacs,longbibliography,floatfix, twocolumn]{revtex4-1}

\usepackage{ragged2e}

\usepackage{color, colortbl}
\definecolor{Gray}{gray}{0.9}
\definecolor{LightCyan}{rgb}{0.88,1,1}
\usepackage{caption}
\usepackage{subcaption}
\usepackage{amsmath}
\usepackage{amssymb}
\usepackage{multirow}

\usepackage[american]{babel}
\usepackage{booktabs}
\usepackage{dcolumn}        
\usepackage[T1]{fontenc}    
\usepackage{graphicx}       
\usepackage[dvipsnames]{xcolor}
\usepackage{hyperref}
\hypersetup{
    colorlinks = true,
    linkcolor= blue,
    citecolor = blue,
    urlcolor = blue,
}
\usepackage[utf8]{inputenc}        
\usepackage{multirow}
\usepackage{txfonts}               
\usepackage{url}
\usepackage{lipsum}
\newcolumntype{d}[1]{D{.}{.}{#1}}
\newcolumntype{v}[1]{D{,}{,\ }{#1}}

\graphicspath{{Figuras/}}

\providecommand{\keywords}[1]
{
  \small	
  \textbf{\textit{Keywords---}} #1
}
\hypersetup{
citecolor = blue,
linkcolor = RoyalPurple,
urlcolor = blue
}

\begin{document}

\title{Interaction in the dark sector: a phenomenological approach}

\author{Z.C. Santana Júnior1$^{1}$} \email{zilmarjunior@hotmail.com.br}
\author{M. O. Costa$^{1}$} \email{marconeoliveiraa@gmail.com}
\author{R.F.L Holanda$^{1}$} \email{holandarfl@fisica.ufrn.br}
\author{R. Silva$^{1,2}$}  \email{raimundosilva@fisica.ufrn.br}

\affiliation{$^{1}$Departamento de F\'{\i}sica, Federal University of Rio Grande do Norte, Natal-RN, 59072-970, Brazil}
\affiliation{$^{2}$Departamento de F\'{\i}sica, Universidade do Estado do Rio Grande do Norte,  Mossor\'o-RN, 59610-210, Brazil}


\date{\today}

\begin{abstract}

    The non-gravitational interaction between the dark components of the Universe could lead to the variation of dark matter energy density standard evolution law. When we assume this scenario, the dark matter energy density follows $\rho_{{dm}}\sim(1+z)^{3 + \epsilon(z)}$ (where $\epsilon(z)=0$ the standard law is recovered). In this paper, we perform a Bayesian analysis to test three parameterizations for $\epsilon(z)$, namely: $\epsilon(z)=\epsilon_0$, $\epsilon(z)=\epsilon_0 + \epsilon_1\frac{z}{1+z}$ and   $\epsilon(z)=\epsilon_0 + \epsilon_1\frac{z(1+z)}{1+z^2}$, where the first one is motivated through the fundamental grounds and the others are on the phenomenological ones. Through the Gaussian process regression, our method uses galaxy cluster gas mass fraction measurements, SNe Ia observations, Cosmic Chronometers, and BAO data. No specific cosmological model is considered. In all possibilities analyzed, the standard evolution law ($\epsilon(z)=0$)  is within $2\sigma$ c.l. The investigated cases generally indicated scenarios of inconclusive or weak evidence toward the simplest model from the Bayesian standpoint.

\end{abstract}

\maketitle

\section{Introduction}\label{sec:intro}

Since the universe's accelerated expansion was discovered about 26 years ago, the observations supporting this dynamic have been accumulating solidly. The most robust evidence comes from observations of supernovae \cite{riess1998observational, perlmutter1999measurements}, cosmic microwave background radiation \cite{aghanim2020planck}, and baryon acoustic oscillations \cite{eisenstein2005detection}. In the context of Einstein's gravitational theory, a fluid with negative pressure, known as dark energy, is necessary to explain this behavior. The most obvious candidate for a dark energy fluid is the cosmological constant, but this leads to one of the significant inconsistencies in physics, namely, the cosmological constant problem \cite{weinberg1989cosmological, padmanabhan2003cosmological, lombriser2023cosmology, sola2022cosmological, sola2013cosmological}. In addition to dark energy, the second most significant component contributing to the universe's energy density is dark matter, a theoretical component that interacts with other components only through gravity (see \cite{2008ARA&A..46..385F,2013PhR...530...87W}).

Together, dark energy (DE) and dark matter (DM) make up about 95\% of the energy of the universe according to the $\Lambda$CDM model – $\Lambda$ for the cosmological constant and CDM for cold dark matter. The $\Lambda$CDM model, or standard model, is well-supported by observational evidence but still has numerous inconsistencies. Besides the cosmological constant problem, other common issues include, e.g., i) Hubble Tension - observations of the Hubble constant ($H_0$) from various sources are incompatible, differing by $4\sigma$ to $6\sigma$ \cite{kamionkowski2022hubble, di2021realm, sola2017h0}, ii)
$\sigma_8$ Discrepancy - discrepancies in the current value of $\sigma_8$ (the present-day amplitude of matter fluctuations in the universe) from different measurements, iii) Coincidence Problem - this problem suggests that we might be living in a particular epoch of the universe where the order of magnitude of the dark energy and dark matter density parameters is equal \cite{zlatev1999quintessence}. In addition to these issues, there is a wide range of problems at different scales (for a comprehensive review, refer to \cite{perivolaropoulos2022challenges, abdalla2022cosmology}).

Many proposals address these issues by considering extensions to the standard model \cite{joyce2015beyond, GmezValent2018, sola2017first}, allowing, for example, for non-gravitational interactions between the dark sector components \cite{wang2016dark}. In this approach, the dark matter and dark energy dominating the current evolution of the universe interact with each other. As their densities coevolve, a natural explanation for the coincidence problem is provided (See \cite{amendola2000coupled} and references therein). The cosmological constant problem is the ground of this scenario (see \cite{carroll2001cosmological, weinberg1989cosmological} for reviews). It is associated with understanding the mechanism that provides small vacuum energy in the same order of magnitude as the present matter density of the universe \cite{perlmutter1999measurements}. Based on the quantum field theory, some approaches have successfully addressed the dynamical cosmological constant \cite{nelson1982scaling, elizalde1994renormalization, elizalde1995gut, bytsenko1994effective, wagoner1970scalar, linde1974lee, polyakov1982phase, cohen1999effective, shapiro2002scaling, shapiro2000scaling, shapiro2003variable, espana2004testing, weinberg1993vacuum,sola2011cosmologies}.

However, this extension must adhere to observational constraints. An energy exchange that significantly influences the evolution law of energy densities can lead to scenarios incompatible with the current observed universe. For example, if the energy density of dark matter evolves much more slowly than predicted by the standard model, it could compromise the formation of structures observed today and the accelerated expansion in the current era of the universe. Additionally, there are reasonable questions regarding thermodynamic limitations. For a more in-depth discussion of this last point, refer to section \ref{secInteracting}. An indication of a non-gravitational interaction between the components of the dark sector is if the energy densities of dark matter and dark energy behave differently from what is expected in the standard model. Therefore, a deviation in the evolution law of the density of a component may indicate unexpected physical processes occurring. Some analyses considering the above issue investigate the possibility of the deviation in the dark matter density evolution law using the gas mass fraction of galaxy clusters and other cosmological probes \cite{holanda2019estimate, bora2021probing, bora2022test}. These previous papers considered $\rho_{{dm}}=\rho_{{dm},0}(1+z)^{3 + \epsilon(z)}$ as $\epsilon(z)=\epsilon_0$ and found observational contraints for $\epsilon_0$. It is worth noting that this parametrization is grounded in fundamental arguments based on the Quantum Field Theory (QFT) (See Refs. \cite{shapiro2002scaling, shapiro2000scaling,GmezValent2018} and references thereby). 
In this paper, a Bayesian analysis to test different parameterizations of a possible deviation of the dark matter energy density standard evolution law  ($\rho_{{dm}}=\rho_{{dm},0}(1+z)^{3 + \epsilon(z)}$) is presented. Three $\epsilon(z)$ possibilities are explored: A constant Deviation, $\epsilon(z)=\epsilon_0$; Chevallier-Polarski-Linder (CPL) parameterization \cite{chevallier2001accelerating, linder2003exploring}, $\epsilon(z)=\epsilon_0 + \epsilon_1\frac{z}{1+z}$; and Barboza-Alcaniz (BA) \cite{barboza2008parametric},  $\epsilon(z)=\epsilon_0 + \epsilon_1\frac{z(1+z)}{1+z^2}$. Such chosen $\epsilon$ functions to explore possible departure are commonly used in the context of the dark sector of cosmology. Galaxy cluster Gas mass fraction data (${f_{\text{gas}}}$), Baryon Acoustic Oscillations measurements (BAO), Type Ia supernovae observations and Hubble parameter from Cosmic Chronometer (CC) values are considered cosmological probes. Specifically, we use Gaussian process regression to reconstruct functions for $D_A(z)$ (using BAO+CC data) and $D_L(z)$ (from SN+CC data), allowing us to estimate distances for the galaxy cluster data. With these two distinct reconstructions, we can perform Bayesian inference to impose constraints on epsilon and compare the competitiveness of the models. In all possibilities, the standard evolution law ($\epsilon(z)=0$)  is within $2\sigma$ c.l.. It is worth emphasizing that the investigated cases indicated scenarios of inconclusive evidence or weak evidence toward the simplest model. Therefore, a model where the deviation would be constant or inconclusive evidence can not significantly differentiate between the models tested here.

Our paper is organized as follows: Section II briefly discusses the interacting model. Section III presents a brief theoretical review of the adopted methodology, while Section IV describes the datasets used in this work. The statistical analysis is described in Section V. In Section VI, we present the results, and in Section VII, we provide a summary of the study's main conclusions.

\section{Interacting Model}\label{secInteracting}

Considering a homogeneous, isotropic, and flat cosmological background described by the Friedmann-Lemaître-Robertson-Walker metric (FLRW) and assuming that the cosmic budget is composed of baryons (b), dark matter (dm), radiation (r), and dark energy (de). The interacting model considers the dark matter and dark energy as interacting fluids with the energy-momentum tensor of the dark sector given by
\begin{equation}\label{coupled-tensor}
	T_{\mu\nu} = T_{\mu\nu}^{{dm}} + T_{\mu\nu}^{{de}}.
\end{equation}

The covariant conservation of energy-momentum tensor, $\nabla_{\mu}T^{\mu\nu} = 0$, leads to
\begin{equation}\label{conservation}
\dot{\rho}_{dm} + 3H\rho_{dm} = -\dot{\rho}_{de} - 3H\rho_{de}(1+\omega) = Q,
\end{equation}
where $\rho_{{dm}}$ and $\rho_{{de}}$ represent the energy density of cold dark matter and dark energy, respectively, while $Q$ is the phenomenological interaction term. Note that $Q > 0$ indicates the dark energy decaying into the dark matter while $Q < 0$ implies the opposite. 

The evolution of the dark components can be found by solving the system of Eqs. (\ref{conservation}). Generally, this can be done by assuming a form for $Q$ \cite{wang2004can, alcaniz2005interpreting, von2019cosmological} or by assuming a relation between the energy densities of the components \cite{cid2019bayesian, von2020unphysical}. Since in the standard description, the dark matter density evolves as $\rho_{dm} \propto (1+z)^{3}$, such a model considers a deviation from the standard evolution characterized by the following function \cite{costa2010cosmological, costa2010coupled}
\begin{equation}\label{dm-evolution}
\rho_{{dm}}=\rho_{{dm},0}(1+z)^{3 + \epsilon(z)},
\end{equation}
where $\rho_{{dm},0}$ is the today dark matter energy density calculated in $z = 0$ and $\epsilon (z)$ is a parametric function that depends of redshift.   

\section{Methodology}\label{sec:modelos}

Galaxy clusters, the largest gravitationally bound structures in the known universe, have the potential to provide a wealth of cosmological information \cite{allen2011cosmological}. In recent years, there has been a growing interest in their observations to conduct cosmological tests \cite{qiu2023cosmology,2023EPJC...83..274B, chaubal2022improving, mantz2022cosmological, corasaniti2021cosmological, wu2021cosmology, holanda2020low, lesci2022amico}. The gas mass fraction is a measurable quantity of these objects, which provides important insights into the baryonic content and the physical processes occurring within galaxy clusters. Due to the scale of its dimensions, it is expected that the mass fraction of galaxy clusters approximately matches the cosmic baryon fraction, $\Omega_b/\Omega_m$, where the subindex $m$ corresponds to total matter \cite{sasaki1996new, allen2008improved, allen2011cosmological, holanda2020low,  mantz2022cosmological}. Thus, it is possible to use estimates of this quantity to constrain cosmological parameters. The $f_{\text{gas}}$ is given by \cite{allen2008improved}: 

\begin{equation}
f_{\text{gas}}(z) = K\gamma \Bigg( \frac{\Omega_b}{\Omega_m} \Bigg) \Bigg[ \frac{D_A^{*}(z)}{D_A(z)}\Bigg]^{3/2} - f_*.
\label{EqGMF1}
\end{equation}
    Here, $D_A$ is the angular diameter distance to the galaxy cluster, $D_A^*$ is the angular diameter distance of the fiducial cosmological model used to infer the gas mass measurement ($H_0 = 70$ km/sec/Mpc, and $\Omega_m = 0.3$ in a flat curvature). $K$ is the calibration constant, which accounts for any inaccuracies in instrument calibration, bias in measured masses due to substructure, bulk motions, and/or non-thermal pressure in the cluster gas \cite{mantz2014cosmology}. $\gamma$ represents the gas depletion factor, a measurement of how much baryonic gas is thermalized within the cluster potential and thereby depleted compared to the cosmic mean \cite{battaglia2013cluster, applegate2016cosmology, holanda2017cosmological}. $\Omega_b$ and $\Omega_m$ are the baryonic and total mass density parameters, respectively. $f_*$ is the fraction of baryonic mass in the form of stars. For the dataset that considers only $r_{2500c}$, we used $f_* = 0$ because the methodology used in the measurements does not account for the luminosity of the stars. Here, we assume that this quantity is independent of the mass of the clusters.

Therefore, taking into account the proposed evolution law, such as $\rho_{dm}=\rho_{dm,0}(1+z)^{3 + \epsilon(z)}$, and that $\Omega_m = \Omega_{dm} + \Omega_b$, the Eq. (\ref{EqGMF1}) can be reformulated in the following manner:

\begin{equation}
(1+z)^{\epsilon(z)} = \Bigg( \frac{\rho_{b,0}}{\rho_{dm,0}} \Bigg) \Bigg[ \Bigg( \frac{K\gamma}{f_{\text{gas}} + f_*} \Bigg)\Bigg( \frac{D_A^*(z)}{D_A(z)} \Bigg)^{3/2} -1 \Bigg].
\label{EqGMF2}
\end{equation}
 We have access to almost all terms on the right-hand side, except for the angular diameter distance for each cluster in the sample. We perform a reconstruction using BAO, CC, and SNe Ia data (see the following subsection) to obtain this quantity for the redshift of each galaxy cluster. In other words, using BAO, CC, and SNe Ia observations, estimating the angular distance for clusters is possible by reconstructing a non-parametric function for $D_A(z)$. 
Regarding $\epsilon(z)$, we test three parameterizations. The models are listed in the Table \ref{TabParametrizations}. The reference model for calculating the Bayes factor was the constant deviation, the simplest model.

\begin{table}
    \centering
    \begin{tabular}{|c|c|}
        \hline
        Model & Parameterization\\
        \hline \hline
        Constant & $\epsilon = \epsilon_0$\\
        \hline
        CPL & $\epsilon = \epsilon_0 + \epsilon_1 \frac{z}{1+z}$\\
        \hline
        BA & $\epsilon = \epsilon_0 + \epsilon_1\frac{z(1+z)}{1+z^2}$ \\
        \hline
    \end{tabular}
    \caption{Parameterizations for the function that represents the possible deviation from the standard model evolution law of the energy density of dark matter used in work}
    \label{TabParametrizations}
\end{table}

Our method uses cluster data to impose constraints on $\rho_{b,0}$ and $\rho_{dm,0}$. We do this by leveraging observations at low redshifts, given that any deviation from the standard model would be negligible for the small redshift range considered. In other words, any constraint obtained with these data is not significantly sensitive to the nature of dark energy \cite{mantz2014cosmology}. Additionally, we can consider $\frac{D_A^*}{D_A} \approx \frac{h}{h_*}$. Hence, Eq. (\ref{EqGMF1}) can be rewritten as:

 \begin{equation}
            \frac{\Omega_b}{\Omega_m}h^{3/2} = \frac{f_{\text{gas}} + f_*}{K\gamma 0.7^{3/2}}.
        \label{EqVincRhob0Rhoc0}
        \end{equation}  

For clusters at $r_{2500c}$, we utilized $5$ data points at $z < 0.16$, whereas for $r_{500c}$, we considered $11$ clusters with $z < 0.09$. By combining Eq. (\ref{EqVincRhob0Rhoc0}) with constraints on the reduced Hubble constant, $h^{\text{SH0ES}} =  0.7324 \pm 0.0174$  \cite{riess20162} or $h^{\text{Planck}} = 67.36 \pm 0.54$ \cite{aghanim2020planck}, and with the value obtained from Big Bang nucleosynthesis for the  $100\Omega_{b,0}h^2$ quantity, $100\Omega_{b,0}h^2 = 2.235 \pm 0.033$ \cite{cooke2018one}, it is possible to derive constraints for $\rho_{b,0}$ and $\rho_{dm,0}$. The results are presented in the Table \ref{Tabrhob0rhoc0}.

\begin{table}
\centering
\begin{tabular}{|@{}c|c|c|}
\hline
Combination & $\rho_{\text{b},0}(\text{x} 10^{-31}gm/cm^3) $ & $\rho_{\text{dm},0}(\text{x} 10^{-31}gm/cm^3) $ \\
\hline
$r_{\text{500c}},K^{\text{CMB}}$ $\gamma^{\text{The300}}$$h^{\text{SH0ES}}$ &  $4.17 \pm 0.207$  &  $13.59 \pm 1.267$  \\
\hline
$r_{\text{500c}},K^{\text{CLASH}}$ $\gamma^{\text{The300}}$$h^{\text{SH0ES}}$ &  $4.17 \pm 0.207$  &  $17.3 \pm 2.022$  \\
\hline
$r_{\text{500c}},K^{\text{CLASH}}$ $\gamma^{\text{FABLE}}$$h^{\text{SH0ES}}$ &  $4.17 \pm 0.207$  &  $17.15 \pm 1.985$  \\
\hline
$r_{\text{500c}},K^{\text{CCCP}}$ $\gamma^{\text{The300}}$$h^{\text{SH0ES}}$ &  $4.17 \pm 0.207$  &  $18.77 \pm 1.597$  \\
\hline
$r_{\text{500c}},K^{\text{CMB}}$ $\gamma^{\text{The300}}$$h^{\text{Planck}}$ &  $4.93 \pm 0.107$  &  $13.61 \pm 1.082$  \\
\hline
$r_{\text{500c}},K^{\text{CLASH}}$ $\gamma^{\text{The300}}$$h^{\text{Planck}}$ &  $4.93 \pm 0.107$  &  $17.37 \pm 1.89$  \\
\hline
$r_{\text{500c}},K^{\text{CLASH}}$ $\gamma^{\text{FABLE}}$$h^{\text{Planck}}$ &  $4.93 \pm 0.107$  &  $17.25 \pm 1.837$  \\
\hline
$r_{\text{500c}},K^{\text{CCCP}}$ $\gamma^{\text{The300}}$$h^{\text{Planck}}$ &  $4.93 \pm 0.107$  &  $18.93 \pm 1.316$  \\
\hline
$r_{\text{2500c}},h^{\text{SH0ES}}$ & $4.17 \pm 0.207$ & $24.49 \pm 3.932$ \\
\hline
$r_{\text{2500c}},h^{\text{Planck}}$ & $4.93 \pm 0.107$ & $25.02 \pm 3.925$ \\
\hline

\end{tabular}
\caption{Estimated values at $\rho_{\text{b},0}$ and $\rho_{\text{dm},0}$ obtained from the Eq.(\ref{EqVincRhob0Rhoc0}) for each combination of $K$, $\gamma$ and $h$.  }
\label{Tabrhob0rhoc0}
\end{table}

\section{Dataset}\label{secdataset}

\subsection{Gas mass fraction}
In this work, we used two sets of galaxy cluster data. For the first one, we utilized a dataset pointed out by \cite{mantz2014cosmology}, consisting of 40 data points of gas mass fraction in galaxy clusters within a redshift range of $0.078 < z < 1.063$. Estimates of this quantity can be derived by observing and analyzing the temperature and density of the X-ray emissions emanating from the intra-cluster medium within galaxy clusters. This dataset consists of gas mass fractions measured in spherical shells within a range of $0.8-1.2r_{2500c}$,  where ``$2500c$'' represents the radius of the cluster where the density of the medium exceeds $2500$ times the critical energy density. For more details on the methodology and the motivation behind this interval, refer to \cite{mantz2014cosmology}.
Furthermore, these samples are confined to the most relaxed clusters to minimize systematic uncertainties and scatter associated with potential departures from hydrostatic equilibrium and spherical symmetry. Very recently, the authors from the Ref. \cite{liu2022dark}, based on a set of $N$-body simulations, investigated the formation histories and properties of dark matter haloes in scenarios where DM-DE interaction occurs and compared with their $\mathrm{\Lambda}$CDM counterparts. Their results showed that dark matter halo formation can be significantly affected by a possible interaction of the two dark components. Therefore, constraints from non-linear structures are indispensable, and the evolution of the gas fraction is a necessary ingredient in the description of the hierarchical growth of clusters.

\begin{figure*}
 \begin{subfigure}{0.45\textwidth}
     \includegraphics[width=\textwidth]{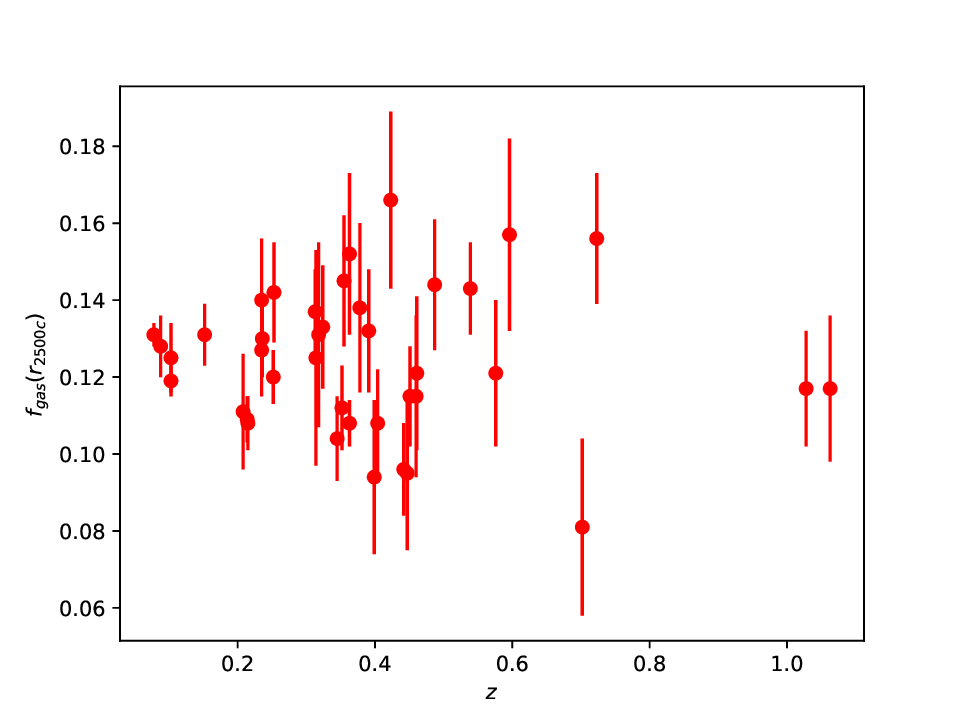}
      \label{fig1:a}
 \end{subfigure}
 \begin{subfigure}{0.45\textwidth}
     \includegraphics[width=\textwidth]{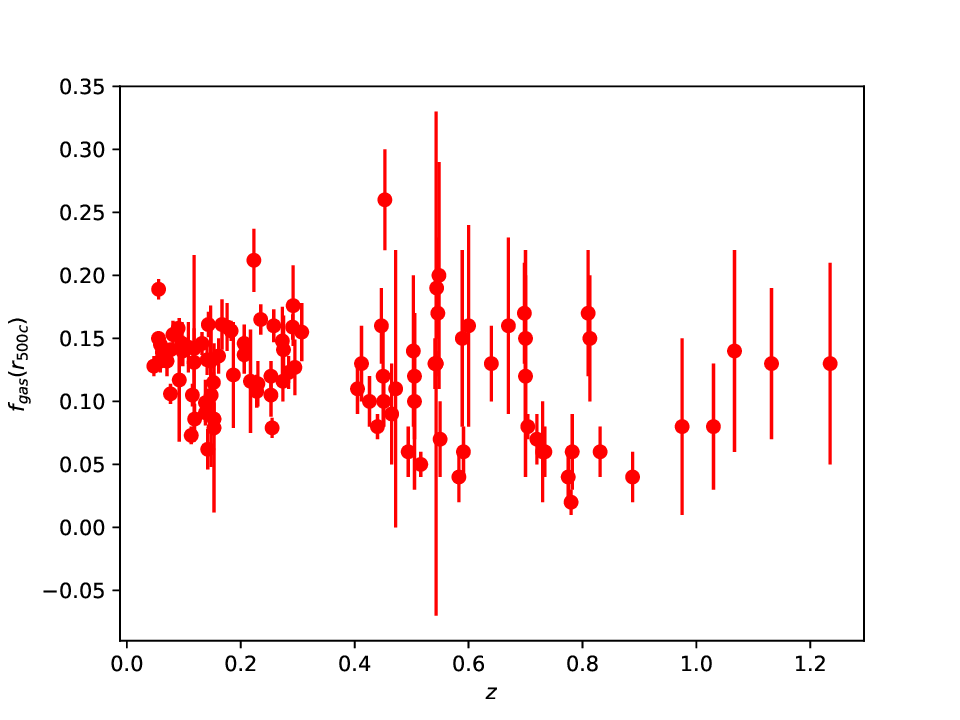}
      \label{fig1:b}
 \end{subfigure}
 \caption{The measurements of the $40$ and $103$ gas mass fraction samples were used in this study.}
 \label{Figdatafgas}
 \end{figure*}

The second set consists of 103 observations on a redshift range of $0.0473 \geq z \geq 1.235$ at $r_{500c}$. This dataset comprises 12 clusters at $z < 0.1$ obtained from X-COP \cite{eckert2019non}; a set of 44 clusters within the range $0.1 \geq z \geq 0.3$ \cite{ettori2010mass}; and observations at high redshifts, consisting of 47 clusters obtained by \cite{ghirardini2017evolution} in the range $0.4 \geq z \geq 1.2$. This dataset consists of measurements for gas mass fraction within $r_{500c}$. In the methodology applied to these observations, the stars mass is also considered, as it contributes to and should be introduced when estimating the gas mass fraction. This set was curated by \cite{corasaniti2021cosmological}. Fig. \ref{Figdatafgas} shows a plot of the two $f_{\text{gas}}$ samples cited.

\subsection{Baryon acoustic oscillations}

We use 18 BAO points within the redshift range of $0.11 \leq z \leq 2.4$, compiled by \cite{staicova2022constraining} to implement the method above. Two different values for the sound horizon distance are considered, namely, $r_d = 136.1 \pm 2.7$ Mpc, obtained through a late-time estimate using H0LiCOW+SN+BAO+SH0ES \cite{arendse2020cosmic}, and $r_d = 147.09 \pm 0.26$ Mpc, based on Planck results \cite{aghanim2020planck}. Thus,  we obtained two datasets of angular diameter distances. However, our methodology requires angular diameter distance measurements for each galaxy cluster. For this, we reconstructed functions for $D_A(z)$ from these datasets using Gaussian process regression, allowing us to estimate this quantity for each galaxy cluster. Fig. \ref{p1fig1-4} (a) and (b) show the angular diameter distance reconstruction function considering both calibrations. We observe that the reconstruction adequately covers the data points and exhibits the expected behavior for the angular distance. This figure shows that the entire gas mass fraction sample is within the redshift range of the BAO and CC data, allowing robust estimates of $D_A(z)$.
	\subsection{Type Ia Supernovae}

   The SNe Ia data include 1048 distance modulus measurements compiled by \cite{scolnic2018complete} (Pantheon Sample)   within redshift range $0.01 \leq z \leq 1.914$. Two values for the supernova absolute magnitude ($M_B$) are considered to obtain luminosity distance data: $M_B=-19.244 \pm 0.037$ (SH0ES team) \cite{camarena2021use}, to get results independent of the cosmological model and $M_B=-19.43 \pm 0.02$, obtained using the Hubble constant estimate by the Planck Collaborations in the context of a $\Lambda$CDM model. The luminosity distances from the reconstruction function for each case can be seen in Fig. \ref{p1fig1-4} (c) and (d). Again, the reconstruction adequately covers the galaxy cluster data points and exhibits the expected behavior for the luminosity distance. As commented earlier, one may obtain $D_A(z)$ by using type Ia Supernovae by considering the cosmic distance duality relation $(D_A(z)=D_L(z)/(1+z)^2)$.

    \subsection{Cosmic Chronometers}
    In addition to BAO and SNe, data from cosmic chronometers were used to reconstruct the $D_A$ and $D_L$ functions. We used $31$ H(z) data obtained from cosmic chronometer (compiled by \cite{li2021testing}) in the redshift $0.07 \leq z \leq 1.965$. This method consists of calculating the derivative of the redshift with respect to cosmic time, $H(z) \simeq -\frac{1}{1+z}\frac{\Delta z}{\Delta t}$ \cite{jimenez2002constraining}. To transform $H(z)$ data to $D_A(z)$ or $D_L(z)$, we solve numerically the comoving distance integral for non-uniformly spaced data using the simple trapezoidal rule

    \begin{equation}
        D_C = \int_0^z\frac{dz'}{H(z')} \approx \frac{c}{2} \sum_{i=1}^N (z_{i+1} - z_i)\Bigg( \frac{1}{H(z_{i+1})} - \frac{1}{H(z_i)} \Bigg).
    \label{EqD_C}
    \end{equation}
    By standard error propagation, the uncertainty associated with the ith bin is 

    \begin{equation}
        s_i = \frac{c}{2}(z_{i+1} - z_i)\Bigg( \frac{\sigma_{H(z_{i+1})}^2}{H^4(z_{i+1})} + \frac{\sigma_{H(z_{i})}^2}{H^4(z_{i})} \Bigg)^{1/2},
    \end{equation}
    so the error of the integral in Eq.(\ref{EqD_C}) from $z = 0$ to $z = z_n$ is $\sigma_n = \sum_{i=1}^n s_i$. For consistency, we added to our sample $H_0^{\text{SH0ES}} = 73.24 \pm 1.74$ or $H_0^{\text{Planck}} = 67.36 \pm 0.54$ to increase the robustness of the analysis. To derive $D_A$ and $D_L$ we considered $\Omega_k = 0$, which translates to $D_A(z) = \frac{1}{1+z}D_C(z)$ and $D_L(z) = (1+z)D_C(z)$. 
    
\section{Statistical Analysis and priors}

\subsection{Bayesian inference}
 In this work, we conducted a model comparison using Bayesian Inference. In this section, we provide a brief introduction to the topic. Bayesian inference is a way to describe the relationship between a model or hypothesis, a set of data, and prior knowledge of that model. The analysis can be summarized by Bayes' theorem, which is expressed as:

\begin{equation}
	\mathcal{P}(\Theta| \mathcal{D}, \mathcal{M}) = \frac{\mathcal{L}(\mathcal{D}|\Theta, \mathcal{M})\pi(\Theta|\mathcal{M})}{\mathcal{E}(\mathcal{D}|\mathcal{M})},
\end{equation}
where $\mathcal{P}(\Theta| \mathcal{D}, \mathcal{M})$ is the posterior distribution, $\mathcal{L}(\mathcal{D}|\Theta, \mathcal{M})$ is the likelihood, $\pi(\Theta|\mathcal{M})$ is the prior distribution, and $\mathcal{E}(\mathcal{D}|\mathcal{M})$ is the Bayesian evidence. The key point of Bayes' theorem is its ability to update our knowledge about the model in the face of new information.

From the perspective of parameter estimation, the evidence, denoted as $\mathcal{E}(\mathcal{D}|\mathcal{M})$, serves merely as a normalization constant. However, when it comes to model comparison, the evidence plays a crucial role in quantifying the model's performance given the data in the parameter space. Thus, we can write the Bayesian evidence as

\begin{equation}
	\mathcal{E}(\mathcal{D}|\mathcal{M}) = \int_\Omega \mathcal{L}(\mathcal{D}|\Theta, \mathcal{M})\mathcal{P}(\Theta |M)d\Theta.
\label{EqEvid}
\end{equation}
Therefore, the evidence represents the average probability value across the parameter space without considering the data.

The most crucial role the evidence plays in Bayesian inference is its role in the model selection problem. It favors models that fit the data well, increasing the average likelihood towards higher values, while penalizing models that have poor fits or are not very predictive, pushing the likelihood towards lower values. As seen from Eq. (\ref{EqEvid}), the evidence favors simpler models, following the principle of Occam's razor.

To perform a direct comparison between the two models, the Bayes factor is used and defined as:

\begin{equation}
	\mathcal{B}_{ij} \equiv \frac{\mathcal{E}_i}{\mathcal{E}_j},
\label{EqBayes}
\end{equation}
where $i$ and $j$ refer to models $\mathcal{M}_i$ and $\mathcal{M}_j$. { In order to implement the Bayesian model selection with the constant $\epsilon(z)$ model as the reference model, let us denote $\mathcal{M}_j$ by $\epsilon=\epsilon_0=const.$. Such a model will be compared to the others. 
}

To analyze the performance of the models according to their evidence, we use Jeffrey's scale shown in Table \ref{TabJeffrey}. This scale is a revision of the original scale, as introduced in \cite{jeffreys1998theory}, made by \cite{trotta2008bayes}. In summary, in the case where $|\text{ln}\mathcal{B}_{ij}| < 1$, there is inconclusive evidence supporting any model. In the scenario where $\text{ln}\mathcal{B}_{ij} < - 1$, model $\mathcal{M}_j$ performs better compared to model $\mathcal{M}_i$. If $\text{ln}\mathcal{B}_{ij} > 1$, model $\mathcal{M}_i$ is favored.

To perform the Bayesian analysis described here, we used PyMultinest \cite{buchner2014x}, a Python module for Multinest \cite{feroz2008multimodal, feroz2009multinest, feroz2013importance}, a tool that employs Importance Nested Sampling to calculate the evidence. This method lets us obtain the posterior distribution, enabling a comprehensive analysis. We used $1,000$ live points for evidence estimation. The priors for $\epsilon_0$ and $\epsilon_1$ in all models were uniform within the interval [-5, 5] for both parameters. The figures were generated using the GetDist \cite{lewis2019getdist} and Matplotlib libraries \cite{hunter2007matplotlib}.

\subsection{Priors}

The values for $K$ and $\gamma$ differ for each dataset due to how $f_{gas}$ are measured. For the $r_{2500c}$ data, we used $K = 0.96 \pm 0.09$ \cite{applegate2016cosmology} and $\gamma = 0.848 \pm 0.085$ \cite{planelles2013baryon}. It is important to point out that Ref. \cite{corasaniti2021cosmological} did not find a significant redshift evolution in the $\gamma$ factor, which will be considered in this analysis. 

For the $r_{500c}$ dataset, we considered $4$ combinations for $K$ and $\gamma$ obtained from the literature. Based on CLASH sample \cite{sereno2015comparing}, we adopted the prior $K^{\text{CLASH}} = 0.78 \pm 0.09$. From the Canadian Cluster Comparison Project (CCCP) \cite{herbonnet2020cccp, hoekstra2015canadian}, we used $K^{\text{CCCP}} = 0.84 \pm 0.04$. As reported in \cite{salvati2018constraints}, a joint analysis of the Planck primary CMB, Planck-SZ number counts, Planck-thermal SZ power spectrum, and BAO provided the value $K^{\text{CMB}} = 0.65 \pm 0.04$. For the depletion factor, we utilized $\gamma^{\text{The300}} = 0.938 \pm 0.041$ inferred from samples of The Three Hundred project \cite{eckert2019non, cui2018three}. Another prior was obtained from the FABLE simulations \cite{henden2020baryon}, where we adopted a constant value of $\gamma^{\text{FABLE}} = 0.931 \pm 0.04$. Finally, the value used for the stellar fraction was $f_* = 0.015 \pm 0.005$ \cite{eckert2019non}.

\begin{table}
    \centering
    \begin{tabular}{|c|c|}
    \hline 
        $\text{ln}\mathcal{B}_{ij}$ & Interpretation \\
    \hline
        Greater than $5$ & Strong evidence for model i\\
        $[2.5, 5]$ & Moderate evidence for model i\\
        $[1, 2.5]$ & Weak evidence for model i\\
        $[-1, 1]$ & Inconclusive \\
        $[-2.5, -1]$ & Weak evidence for model j\\
        $[-5, -2.5]$ & Moderate evidence for model j\\
        Less than $-5$ & Strong evidence for model j\\
    \hline
    \end{tabular}
    \caption{The table shows the values of the Bayes factor and their corresponding interpretation.}
    \label{TabJeffrey}
\end{table}

\section{Results}\label{secresults}

\begin{figure*}
    \centering
    \begin{subfigure}[b]{0.45\textwidth}
        \centering
        \includegraphics[width=\textwidth]{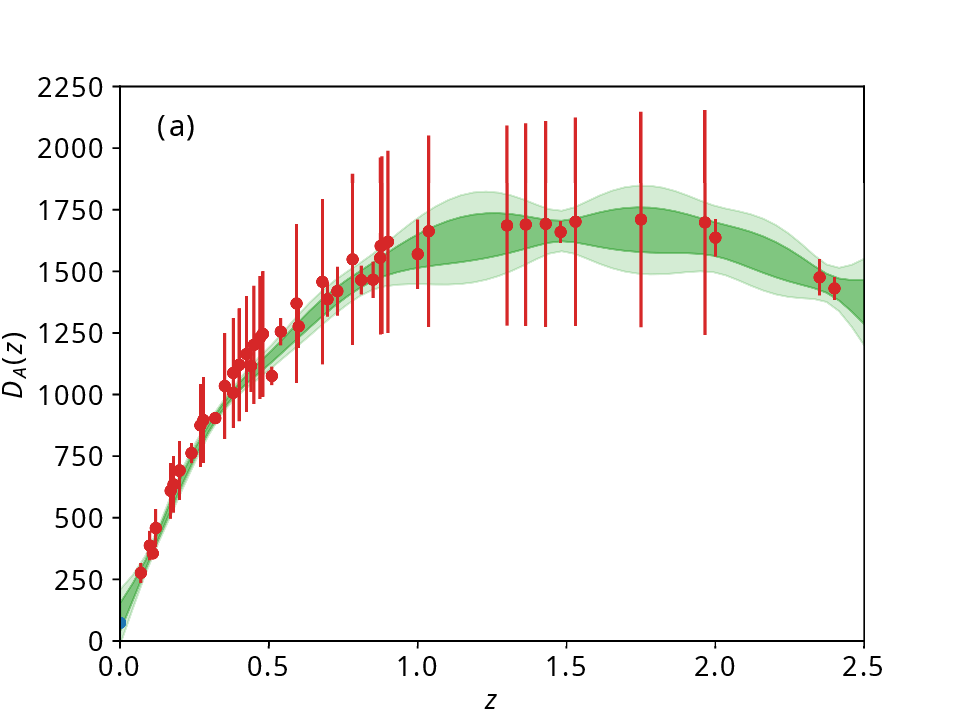}
        \label{fig:comb1}
    \end{subfigure}
    \hfill
    \begin{subfigure}[b]{0.45\textwidth}
        \centering
        \includegraphics[width=\textwidth]{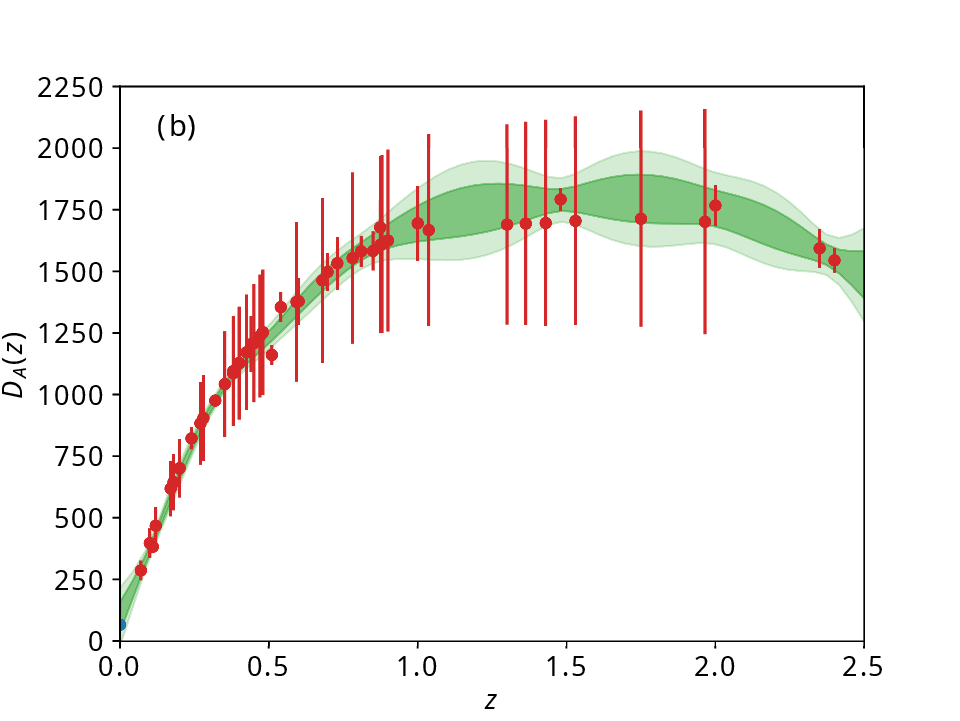}
        \label{fig:comb2}
    \end{subfigure}

    \vspace{0.5cm}

    \begin{subfigure}[b]{0.45\textwidth}
        \centering
        \includegraphics[width=\textwidth]{ 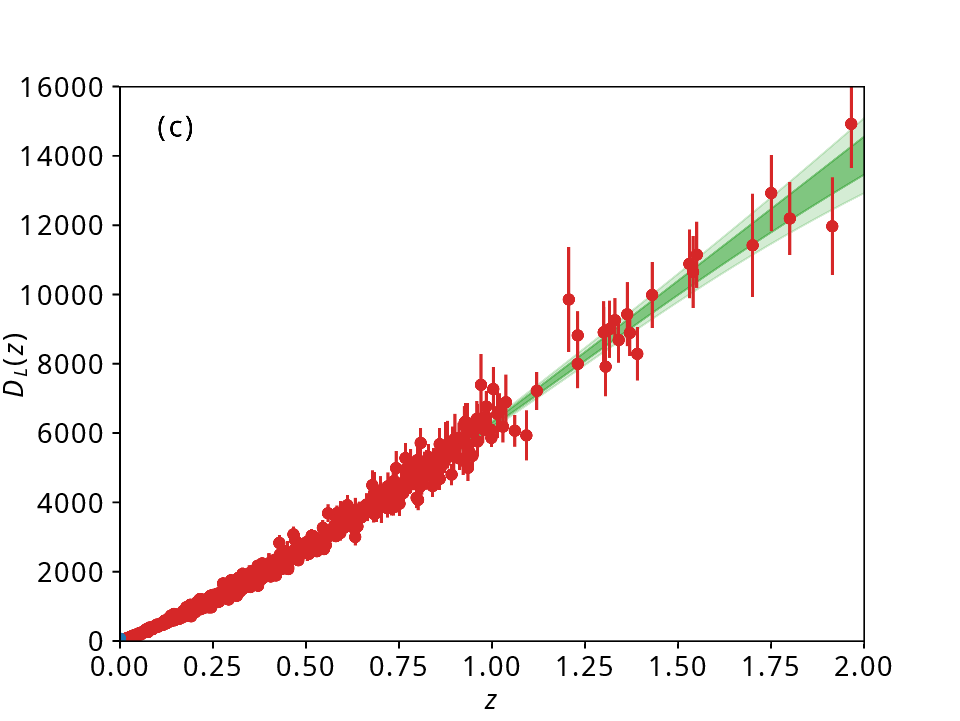}
        \label{fig:comb3}
    \end{subfigure}
    \hfill
    \begin{subfigure}[b]{0.45\textwidth}
        \centering
        \includegraphics[width=\textwidth]{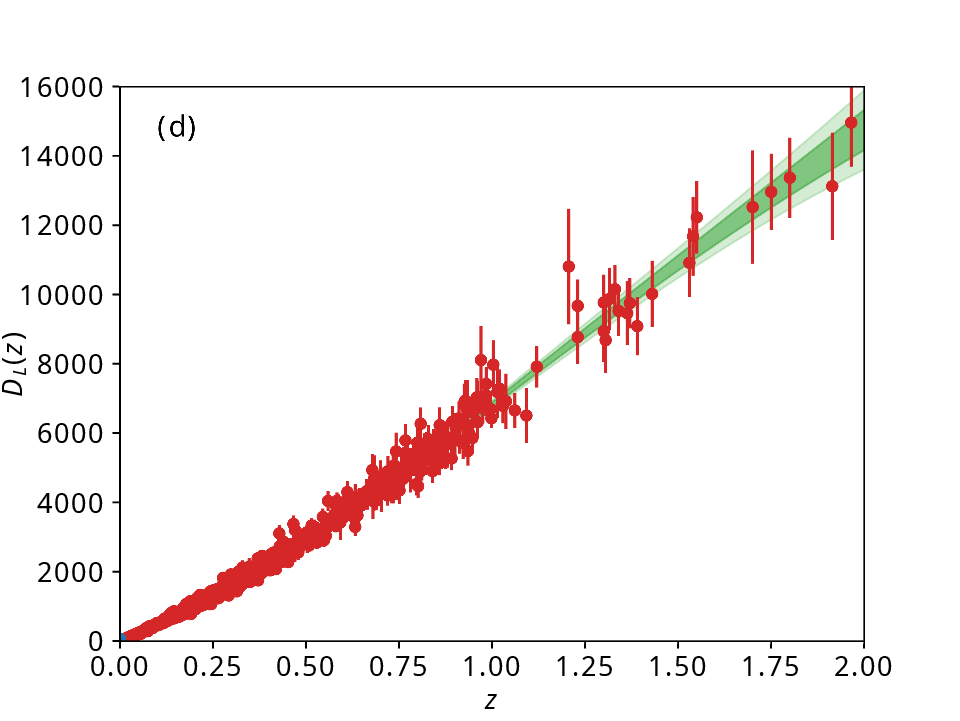}
        \label{fig:comb4}
    \end{subfigure}
    \caption{Panel with the four reconstructions for the distances. Figs. (a) and (b) show the functions for the angular diameter distance for the cases $h^{\text{Planck}}$ and $h^{\text{SH0ES}}$. Figs. (c) and (d) show the functions for the luminosity distance for $h^{\text{Planck}}$ and $h^{\text{SH0ES}}$, respectively. The red points represent the data used for Gaussian process regression, while the dark green and light green areas represent the $68\%$ and $95\%$ confidence levels, respectively. The blue dots represent the values of $H_0$ used in the CC data consistent with the $h$ of each case.}
    \label{p1fig1-4}
\end{figure*}

The reconstructions of $D_A(z)$ and $D_L(z)$ functions using BAO+CC and SN+CC data are presented in Fig. \ref{p1fig1-4}. The behavior is as expected for a flat FLRW universe, covering the entire redshift range of the used $f_{\text{gas}}$ data. Hence, it is possible to robustly estimate the angular diameter distance of galaxy clusters.

\begin{figure*}
 \begin{subfigure}{0.24\textwidth}
 \caption{}
     \includegraphics[width=\textwidth]{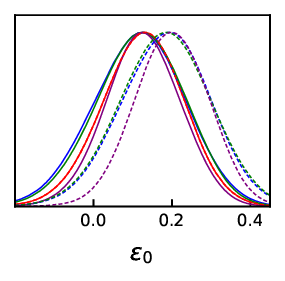}
      \label{fig2:a}
 \end{subfigure}
 \hspace{5cm}
  \begin{subfigure}{0.24\textwidth}
     \includegraphics[width=\textwidth]{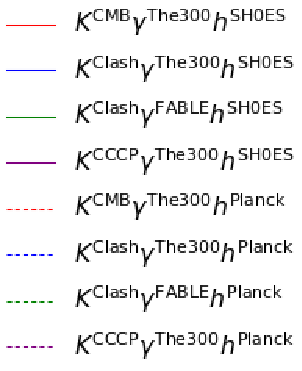}

 \end{subfigure}
 \newline
 \hfill
 \begin{subfigure}{0.48\textwidth}
 \caption{}
     \includegraphics[width=\textwidth]{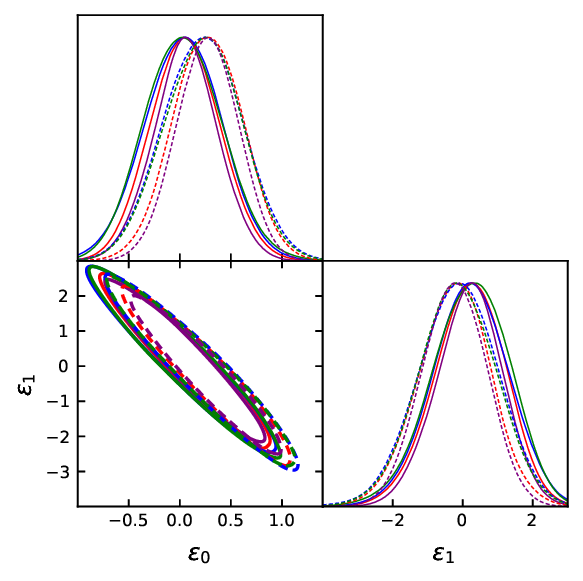}
      \label{fig2:b}
 \end{subfigure}
 \begin{subfigure}{0.48\textwidth}
 \caption{}
     \includegraphics[width=\textwidth]{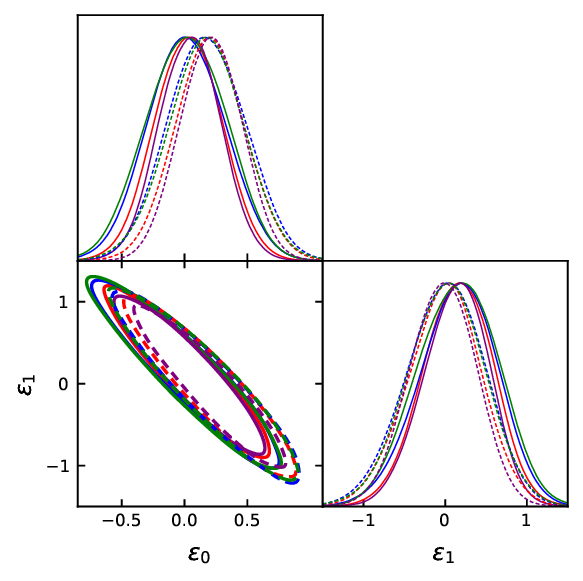}
      \label{fig2:c}
 \end{subfigure}
 \hfill

  \caption{Constraints on the parameters obtained using BAO and the cluster dataset at $r_{500c}$. The captions a), b), and c) refer to constant, CPL, and BA parameterizations, respectively.}
     \label{P2-R500-BAO}
\end{figure*}

\begin{figure*}
 \begin{subfigure}{0.24\textwidth}
 \caption{}
     \includegraphics[width=\textwidth]{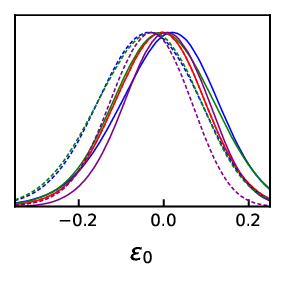}
      \label{fig3:a}
 \end{subfigure}
 \hspace{5cm}
  \begin{subfigure}{0.24\textwidth}
     \includegraphics[width=\textwidth]{legenda.eps}

 \end{subfigure}
 \newline
 \hfill
 \begin{subfigure}{0.48\textwidth}
 \caption{}
     \includegraphics[width=\textwidth]{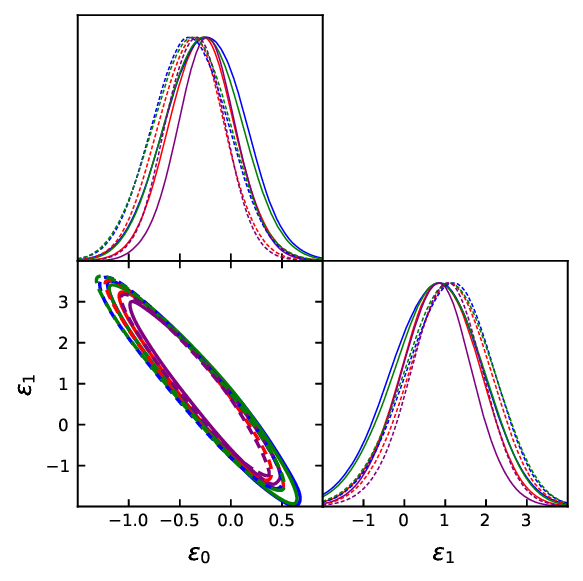}
      \label{fig3:b}
 \end{subfigure}
 \begin{subfigure}{0.48\textwidth}
 \caption{}
     \includegraphics[width=\textwidth]{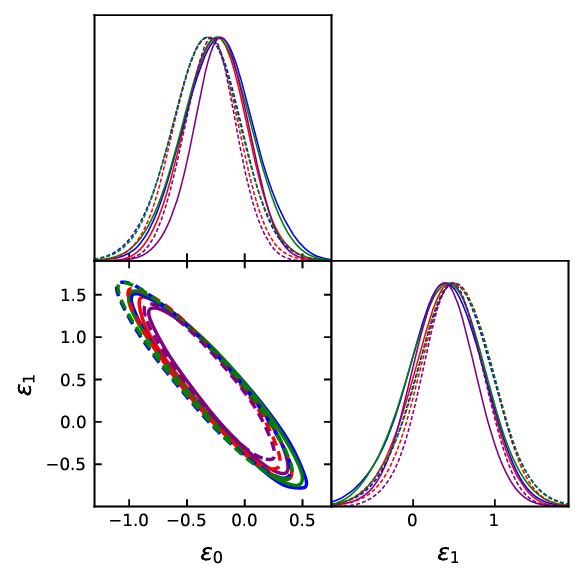}
      \label{fig3:c}
 \end{subfigure}
 \hfill

  \caption{Constraints on the parameters obtained using SN and the cluster dataset at $r_{500c}$. The captions a), b), and c) refer to constant, CPL, and BA parameterizations, respectively.}
     \label{P3-R500-SN}
\end{figure*}

\begin{figure*}
\begin{center}
 \begin{subfigure}{0.24\textwidth}
 \caption{}
     \includegraphics[width=\textwidth]{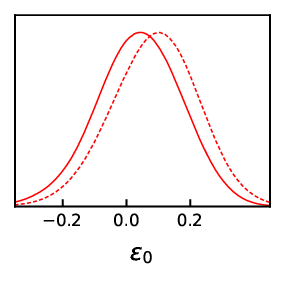}
      \label{fig4:a}
 \end{subfigure}
\begin{subfigure}{0.34\textwidth}
 \caption{}   
     \includegraphics[width=\textwidth]{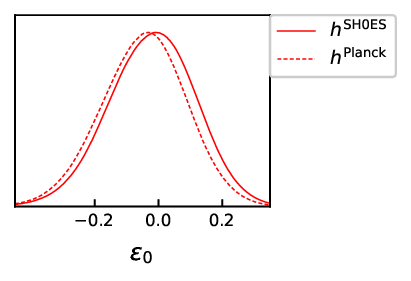}
     \label{P4fig3}
 \end{subfigure}
 \end{center}
 \hfill
 \begin{subfigure}{0.48\textwidth}
 \caption{}
     \includegraphics[width=\textwidth]{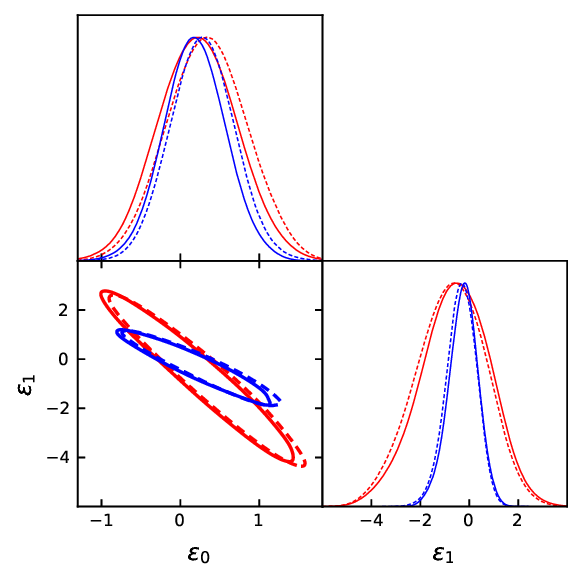}
      \label{fig4:c}
 \end{subfigure}
 \begin{subfigure}{0.48\textwidth}
 \caption{}
     \includegraphics[width=\textwidth]{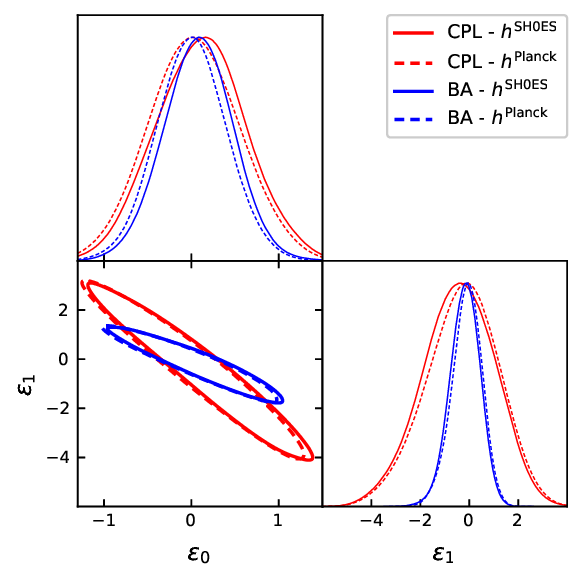}
      \label{fig4:d}
 \end{subfigure}
 \hfill

  \caption{Constraints on the parameters obtained using the cluster dataset at $r_{2500c}$. The captions a) and b) refer to the constant model considering the BAO and SN datasets in this order. The captions c) and d) refer to the constraints obtained considering BAO and SN datasets and comprise the results of the models CPL and BA.}
     \label{P4-R2500-BAOSN}
\end{figure*}

Regarding Bayesian inference, the results are presented in Tables \ref{TabR500BAO}, \ref{Tabr500SN}, \ref{TabR2500BAO} and \ref{TabR2500SN}. The Figs. \ref{P2-R500-BAO} and \ref{P3-R500-SN} show the obtained contraints distributions to all combinations for $r_{\text{500c}}$ using BAO+CC and SN+CC data respectively. The distributions of the analyses made with $r_{2500c}$ galaxy cluster data are presented in Fig. \ref{P4-R2500-BAOSN}, showing the results of BAO+CC and SN+CC methodologies simultaneously. The results show an agreement within $2\sigma$ of the standard model.

Finally, when comparing the adopted models, we calculated the natural logarithm of the evidence (Eq. (\ref{EqEvid})) and the natural logarithm of the Bayes factor (Eq. (\ref{EqBayes})). We can analyze the performance of each adopted model using Jeffreys' scale, as presented in Table \ref{TabJeffrey}. Overall, we can see two scenarios: almost all cases suggest weak support for the simpler model, with a few exceptions with inconclusive evidence. It's important to note that all cases agree with each other and the standard scenario. Thus, this preference for the constant model is due to its simplicity, having one degree of freedom less than the tested models. Furthermore, it's worth mentioning that the uncertainties in the data do not allow for a robust differentiation among the proposed scenarios.

\begin{table*}[htbp]
    \centering
    \begin{tabular}{|c|c|c|c|c|c|c|}
        \hline
        & Model & $\epsilon_0$ & $\epsilon_1$ & ln$\mathcal{E}$ & ln$\mathcal{B}_{i1}$ & Interpretation \\
        \hline
        \multirow{5}{*}{$K^{\text{CMB}}\gamma^{\text{The300}}h^{\text{SH0ES}}$} & Constant & $0.118 \pm 0.104$ & & -119.001 $\pm$ 0.007 & & \\
        \cline{2-7}
        & CPL & $0.0391 \pm 0.339$ & $0.233 \pm 1.012$ & -120.185 $\pm$ 0.006 & -1.184 $\pm$ 0.009 & Weak evidence against \\
        \cline{2-7}
        & BA & $0.0215 \pm 0.269$ & $0.163 \pm 0.431$ & -121.025 $\pm$ 0.006 & -2.024 $\pm$ 0.009 & Weak evidence against \\
        \hline
        \multirow{5}{*}{$K^{\text{Clash}}\gamma^{\text{The300}}h^{\text{SH0ES}}$}  & Constant & $0.117 \pm 0.113$ & & -104.392 $\pm$ 0.007 & & \\
        \cline{2-7}
        & CPL & $0.0448 \pm 0.381$ & $0.199 \pm 1.103$ & -105.485 $\pm$ 0.006 & -1.09 $\pm$ 0.009 & Weak evidence against \\
        \cline{2-7}
        & BA & $0.015 \pm 0.302$ & $0.155 \pm 0.464$ & -106.302 $\pm$ 0.006 & -1.91 $\pm$ 0.009 & Weak evidence against \\
        \hline
        \multirow{5}{*}{$K^{\text{Clash}}\gamma^{\text{FABLE}}h^{\text{SH0ES}}$}  & Constant & $0.110 \pm 0.114$ & & -104.222 $\pm$ 0.008 & & \\
        \cline{2-7}
        & CPL & $0.019 \pm 0.382$ & $0.251 \pm 1.104$ & -105.413 $\pm$ 0.007 & -1.191 $\pm$ 0.010 & Weak evidence against \\
        \cline{2-7}
        & BA & $0.007 \pm 0.302$ & $0.161 \pm 0.464$ & -106.28 $\pm$ 0.005 & -2.058 $\pm$ 0.009 & Weak evidence against \\
        \hline
        \multirow{5}{*}{$K^{\text{CCCP}}\gamma^{\text{The300}}h^{\text{SH0ES}}$} & Constant & $0.126 \pm 0.096$ & & -118.628 $\pm$ 0.008 & & \\
        \cline{2-7}
        & CPL & $0.059 \pm 0.303$ & $0.195 \pm 0.913$ & -120.011 $\pm$ 0.006 & -1.383 $\pm$ 0.010 & Weak evidence against \\
        \cline{2-7}
        & BA & $0.0417 \pm 0.247$ & $0.16 \pm 0.398$ & -120.858 $\pm$ 0.006 & -2.230 $\pm$ 0.01 & Weak evidence against \\
        \hline
        \multirow{5}{*}{$K^{\text{CMB}}\gamma^{\text{The300}}h^{\text{Planck}}$} & Constant & $0.202 \pm 0.101$ & & -127.047 $\pm$ 0.008 & & \\
        \cline{2-7}
        & CPL & $0.264 \pm 0.345$ & $-0.221 \pm 1.039$ & -128.394 $\pm$ 0.007 & -1.347 $\pm$ 0.010 & Weak evidence against \\
        \cline{2-7}
        & CPL & $0.197 \pm 0.266$ & $-0.007 \pm 0.437$ & -129.304 $\pm$ 0.006 & -2.257 $\pm$ 0.01 & Weak evidence against \\
        \hline
        \multirow{5}{*}{$K^{\text{Clash}}\gamma^{\text{The300}}h^{\text{Planck}}$}  & Constant & $0.175 \pm 0.114$ & & -109.128 $\pm$ 0.007 & & \\
        \cline{2-7}
        & CPL & $0.212 \pm 0.377$ & $-0.129 \pm 1.1$ & -110.369 $\pm$ 0.008 & -1.240 $\pm$ 0.011 & Weak evidence against \\
        \cline{2-7}
        & BA & $0.163 \pm 0.303$ & $0.012 \pm 0.472$ & -111.297 $\pm$ 0.005 & -2.168 $\pm$ 0.009 & Weak evidence against \\
        \hline
        \multirow{5}{*}{$K^{\text{Clash}}\gamma^{\text{FABLE}}h^{\text{Planck}}$}  & Constant & $0.183 \pm 0.113$ & & -109.157 $\pm$ 0.009 & & \\
        \cline{2-7}
        & CPL & $0.224 \pm 0.382$ & $-0.167 \pm 1.116$ & -110.456 $\pm$ 0.007 & -1.299 $\pm$ 0.011 & Weak evidence against \\
        \cline{2-7}
        & BA & $0.163 \pm 0.315$ & $0.017 \pm 0.490$ & -111.407 $\pm$ 0.006 & -2.250 $\pm$ 0.010 & Weak evidence against \\
        \hline
        \multirow{5}{*}{$K^{\text{CCCP}}\gamma^{\text{The300}}h^{\text{Planck}}$}  & Constant & $0.191 \pm 0.091$ & & -128.045 $\pm$ 0.008 & & \\
        \cline{2-7}
        & CPL & $0.254 \pm 0.3$ & $-0.209 \pm 0.922$ & -129.531 $\pm$ 0.007 & -1.486 $\pm$ 0.011 & Weak evidence against \\
        \cline{2-7}
        & BA & $0.195 \pm 0.242$ & $-0.015 \pm 0.399$ & -130.478 $\pm$ 0.006 & -2.433 $\pm$ 0.01 & Weak evidence against \\
        \hline
    \end{tabular}
    \vspace{20pt} 
    \caption{Summary of models with corresponding values of $\epsilon_0$, $\epsilon_1$, Evidence, and Bayes Factor for the $r_{500c}$ dataset. The priors of $K$, $\gamma$, and $h$ are designated in the table. This analysis used BAO data to reconstruct $D_A(z)$.}
    \label{TabR500BAO}
\end{table*}

\begin{table*}[htbp]
    \centering
    \begin{tabular}{|c|c|c|c|c|c|c|}
        \hline
        & Model & $\epsilon_0$ & $\epsilon_1$ & ln$\mathcal{E}$ & ln$\mathcal{B}_{i1}$ & Interpretation \\
        \hline
        \multirow{5}{*}{$K^{\text{CMB}}\gamma^{\text{The300}}h^{\text{SH0ES}}$}  & Constant & $-0.004 \pm 0.105$ & & -117.843 $\pm$ 0.007 & & \\
        \cline{2-7}
        & CPL & $-0.267 \pm 0.326$ & $0.829 \pm 0.978$ & -118.731 $\pm$ 0.005 & -0.888 $\pm$ 0.009 & Inconclusive \\
        \cline{2-7}
        & BA & $-0.242 \pm 0.262$ & $0.414 \pm 0.416$ & -119.456 $\pm$ 0.006 & -1.613 $\pm$ 0.009 & Weak evidence against \\
        \hline
        \multirow{5}{*}{$K^{\text{Clash}}\gamma^{\text{The300}}h^{\text{SH0ES}}$}  & Constant & $0.008 \pm 0.112$ & & -102.283 $\pm$ 0.008 & & \\
        \cline{2-7}
        & CPL & $-0.247 \pm 0.374$ & $0.761 \pm 1.086$ & -103.270 $\pm$ 0.006 & -0.987 $\pm$ 0.010 & Inconclusive \\
        \cline{2-7}
        & BA & $-0.23 \pm 0.307$ & $0.393 \pm 0.473$ & -104.068 $\pm$ 0.006 & -1.785 $\pm$ 0.01 & Weak evidence against \\
        \hline
        \multirow{5}{*}{$K^{\text{Clash}}\gamma^{\text{FABLE}}h^{\text{SH0ES}}$}  & Constant & $0.004 \pm 0.115$ & & -102.236 $\pm$ 0.008 & & \\
        \cline{2-7}
        & CPL & $-0.269 \pm 0.379$ & $0.818 \pm 1.095$ & -103.194 $\pm$ 0.006 & -0.958 $\pm$ 0.01 & Inconclusive \\
        \cline{2-7}
        & BA & $-0.233 \pm 0.302$ & $0.393 \pm 0.471$ & -103.971 $\pm$ 0.006 & -1.735 $\pm$ 0.01 & Weak evidence against \\
        \hline
        \multirow{5}{*}{$K^{\text{CCCP}}\gamma^{\text{The300}}h^{\text{SH0ES}}$}  & Constant & $0.011 \pm 0.094$ & & -118.019 $\pm$ 0.007 & & \\
        \cline{2-7}
        & CPL & $-0.21 \pm 0.296$ & $0.729 \pm 0.891$ & -119.067 $\pm$ 0.007 & -1.048 $\pm$ 0.01 & Weak evidence against \\
        \cline{2-7}
        & BA & $-0.198 \pm 0.244$ & $0.365 \pm 0.4$ & -119.859 $\pm$ 0.006 & -1.84 $\pm$ 0.009 & Weak evidence against \\
        \hline
        \multirow{5}{*}{$K^{\text{CMB}}\gamma^{\text{The300}}h^{\text{Planck}}$}  & Constant & $-0.061 \pm 0.104$ & & -124.326 $\pm$ 0.008 & & \\
        \cline{2-7}
        & CPL & $-0.416 \pm 0.325$ & $1.138 \pm 0.985$ & -125.079 $\pm$ 0.006 & -0.753 $\pm$ 0.01 & Inconclusive \\
        \cline{2-7}
        & BA & $-0.374 \pm 0.259$ & $0.559 \pm 0.422$ & -125.770 $\pm$ 0.005 & -1.444 $\pm$ 0.009 & Weak evidence against \\
        \hline
        \multirow{5}{*}{$K^{\text{Clash}}\gamma^{\text{The300}}h^{\text{Planck}}$} & Constant & $-0.047 \pm 0.116$ & & -104.957 $\pm$ 0.007 & & \\
        \cline{2-7}
        & CPL & $-0.42 \pm 0.377$ & $1.133 \pm 1.09$ & -105.722 $\pm$ 0.006 & -0.765 $\pm$ 0.009 & Inconclusive \\
        \cline{2-7}
        & BA & $-0.372 \pm 0.305$ & $0.539 \pm 0.472$ & -106.459 $\pm$ 0.006 & -1.501 $\pm$ 0.009 & Weak evidence against \\
         \hline
        \multirow{5}{*}{$K^{\text{Clash}}\gamma^{\text{FABLE}}h^{\text{Planck}}$}  & Constant & $-0.035 \pm 0.118$ & & -105.473 $\pm$ 0.009 & & \\
        \cline{2-7}
        & CPL & $-0.378 \pm 0.365$ & $1.032 \pm 1.059$ & -106.362 $\pm$ 0.006 & -0.89 $\pm$ 0.011 & Inconclusive \\
        \cline{2-7}
        & BA & $-0.331 \pm 0.302$ & $0.496 \pm 0.470$ & -107.095 $\pm$ 0.006 & -1.622 $\pm$ 0.010 & Weak evidence against \\
        \hline
        \multirow{5}{*}{$K^{\text{CCCP}}\gamma^{\text{The300}}h^{\text{Planck}}$}  & Constant & $-0.040 \pm 0.094$ & & -125.436 $\pm$ 0.007 & & \\
        \cline{2-7}
        & CPL & $-0.353 \pm 0.289$ & $1.013 \pm 0.882$ & -126.281 $\pm$ 0.006 & -0.845 $\pm$ 0.009 & Inconclusive \\
        \cline{2-7}
        & BA & $-0.313 \pm 0.233$ & $0.495 \pm 0.381$ & -126.968 $\pm$ 0.006 & -1.532 $\pm$ 0.009 & Weak evidence against \\
        \hline
    \end{tabular}
    \vspace{20pt}
    \caption{Summary of models with corresponding values of $\epsilon_0$, $\epsilon_1$, Evidence, and Bayes Factor for the $r_{500c}$ dataset. The priors of $K$, $\gamma$, and $h$ are designated in the table. This analysis used SN data to reconstruct $D_L(z)$.}
    \label{Tabr500SN}
\end{table*}

\begin{table*}[htbp]
    \centering
    \begin{tabular}{|c|c|c|c|c|c|c|}
        \hline
        & Model & $\epsilon_0$ & $\epsilon_1$ & ln$\mathcal{E}$ & ln$\mathcal{B}_{i1}$ & Interpretation \\
        \hline
        \multirow{5}{*}{$h^{\text{SH0ES}}$}  & Constant & $0.043 \pm 0.133$ & & -21.399 $\pm$ 0.006 & & \\
        \cline{2-7}
        & CPL & $0.24 \pm 0.499$ & $-0.634 \pm 1.416$ & -22.204 $\pm$ 0.007 & -0.805 $\pm$ 0.009 & Inconclusive \\
        \cline{2-7}
        & BA & $0.202 \pm 0.392$ & $-0.276 \pm 0.600$ & -23.091 $\pm$ 0.006 & -1.692 $\pm$ 0.008 & Weak evidence against \\
        \hline
        \multirow{5}{*}{$h^{\text{Planck}}$}  & Constant & $0.081 \pm 0.137$ & & -21.864 $\pm$ 0.007 & & \\
        \cline{2-7}
        & CPL & $0.329 \pm 0.493$ & $-0.758 \pm 1.403$ & -22.848 $\pm$ 0.006 & -0.984 $\pm$ 0.009 & Inconclusive \\
        \cline{2-7}
        & BA & $0.273 \pm 0.407$ & $-0.332 \pm 0.628$ & -23.746 $\pm$ 0.006 & -1.882 $\pm$ 0.009 & Weak evidence against \\
        \hline
    \end{tabular}
    \vspace{20pt} 
    \caption{Summary of the results obtained for the parameters, evidence, Bayes factor, and their respective interpretations for each model concerning the $r_{2500c}$ data. We use the Gaussian process to reconstruct $D_A(z)$ from BAO data.}
    \label{TabR2500BAO}
\end{table*}

\begin{table*}[htbp]
    \centering
    \begin{tabular}{|c|c|c|c|c|c|c|}
        \hline
        & Model & $\epsilon_0$ & $\epsilon_1$ & ln$\mathcal{E}$ & ln$\mathcal{B}_{i1}$ & Interpretation \\
        \hline
        \multirow{5}{*}{$h^{\text{SH0ES}}$}  & Constant & $-0.019 \pm 0.136$ & & -20.950 $\pm$ 0.008 & & \\
        \cline{2-7}
        & CPL & $0.116 \pm 0.507$ & $-0.434 \pm 1.434$ & -21.827 $\pm$ 0.007 & -0.88 $\pm$ 0.01 & Inconclusive \\
        \cline{2-7}
        & BA & $0.077 \pm 0.399$ & $-0.179 \pm 0.621$ & -22.759 $\pm$ 0.006 & -1.81 $\pm$ 0.01 & Weak evidence against \\
        \hline
        \multirow{5}{*}{$h^{\text{Planck}}$}  & Constant & $-0.059 \pm 0.133$ & & -20.812 $\pm$ 0.008 & & \\
        \cline{2-7}
        & CPL & $0.027 \pm 0.499$ & $-0.27 \pm 1.43$ & -21.878 $\pm$ 0.007 & -1.066 $\pm$ 0.009 & Weak evidence against \\
        \cline{2-7}
        & BA & $0.008 \pm 0.403$ & $-0.123 \pm 0.625$ & -22.812 $\pm$ 0.006 & -2.0 $\pm$ 0.009 & Weak evidence against \\
        \hline
    \end{tabular}
    \vspace{20pt} 
    \caption{Summary of the results obtained for the parameters, evidence, Bayes factor, and their respective interpretations for each model concerning the $r_{2500c}$ data. We use the Gaussian process to reconstruct $D_L(z)$ from SN data.}
    \label{TabR2500SN}
\end{table*}

Table \ref{tabResultsComparacao} compares the constraints obtained for $\epsilon_0$ of constant model in this work and those from other studies using the same methodological approach. It is possible to see that within the uncertainty level, there is an agreement among all results. However, we can make a noteworthy conclusion. The constraints obtained for $\epsilon_0$ for the $r_{500}$ gas mass fraction data and estimated from the reconstruction using supernova data are smaller than those from BAO. The variation found out to $\epsilon_0$ is primarily associated with the different values for the parameters used, i.e., $K$, $\gamma$, $h$ and their respective uncertainties, as well as the data characteristics ($f_{\text{gas}}$). Nevertheless, the values generally agree with each other within 1$\sigma$. The most notable cases are the supernova data, which provide a more significant constraint for the scenario where $\epsilon_0 = 0$, representing a case without interaction. In the scenarios for the $r_{2500}$ $f_{gas}$ dataset, the analyzed cases show no significant difference.
Moreover, it is worth stressing that in the phenomenological ground and using CMB, SN Ia, and the Hubble evolution $H(z)$, the authors of Ref. \cite{Costa2009} found the $\epsilon_0\sim 0.0$, which is equivalent to $\Lambda$CDM model. Yet, based on the fundamental ground (See Refs. \cite{shapiro2002scaling, shapiro2000scaling,GmezValent2018, sola2023running} and references thereby), some authors addressed the interaction among the dark components in the Quantum Field Theory (QFT) context. As a result, they proposed the so-called running vacuum models (RVMs). In this scenario, the value $\nu\sim10^{-3}$, here mapped by $\epsilon_0$, was estimated through Grand Unified Theory (GUT) and using the cosmological data SN Ia+BAO+H(z)+LSS+CMB.

\begin{table}[]
    \centering
    \begin{tabular}{|c|c|c|}
        \hline
         Dataset & $\epsilon_0$ & Reference  \\ \hline
         $f_{gas}$ + SN & $0.13 \pm 0.235$  & \cite{holanda2019estimate} \\ \hline
         $f_{gas}$ + Strong Gravitational Lensing & $-0.088 \pm 0.11 $ & \cite{bora2021probing} \\ \hline
         $f_{gas}$ + Cosmic Chronometers & $-0.234^{+0.159}_{-0.171} $ & \cite{bora2022test} \\ \hline
         $f_{gas,r500}$ + BAO ($K^{\text{CMB}}\gamma^{\text{The300}}h^{\text{SH0ES}}$) & $0.118 \pm 0.104$ & This work \\ \hline
         $f_{gas, r500}$ + BAO ($K^{\text{Clash}}\gamma^{\text{The300}}h^{\text{SH0ES}}$) & $0.117 \pm 0.113$ & This work \\ \hline
         $f_{gas, r500}$ + BAO ($K^{\text{Clash}}\gamma^{\text{FABLE}}h^{\text{SH0ES}}$) & $0.110 \pm 0.114$ & This work \\ \hline
         $f_{gas, r500}$ + BAO ($K^{\text{CCCP}}\gamma^{\text{The300}}h^{\text{SH0ES}}$) &  $0.126 \pm 0.096$ & This work \\ \hline
         $f_{gas, r500}$ + BAO ($K^{\text{CMB}}\gamma^{\text{The300}}h^{\text{Planck}}$) &  $0.202 \pm 0.101$ & This work \\ \hline
         $f_{gas, r500}$ + BAO ($K^{\text{Clash}}\gamma^{\text{The300}}h^{\text{Planck}}$) & $0.175 \pm 0.114$ & This work \\ \hline
         $f_{gas, r500}$ + BAO ($K^{\text{Clash}}\gamma^{\text{FABLE}}h^{\text{Planck}}$) & $0.183 \pm 0.113$ & This work \\ \hline
         $f_{gas, r500}$ + BAO ($K^{\text{CCCP}}\gamma^{\text{The300}}h^{\text{Planck}}$) & $0.191 \pm 0.091$  & This work \\ \hline
         $f_{gas, r500}$ + SN ($K^{\text{CMB}}\gamma^{\text{The300}}h^{\text{SH0ES}}$) & $-0.004 \pm 0.105$ & This work \\ \hline
         $f_{gas, r500}$ + SN ($K^{\text{Clash}}\gamma^{\text{The300}}h^{\text{SH0ES}}$) & $0.008 \pm 0.112$  & This work \\ \hline
         $f_{gas, r500}$ + SN ($K^{\text{Clash}}\gamma^{\text{FABLE}}h^{\text{SH0ES}}$) & $0.004 \pm 0.115$ & This work \\ \hline
         $f_{gas, r500}$ + SN ($K^{\text{CCCP}}\gamma^{\text{The300}}h^{\text{SH0ES}}$) & $0.011 \pm 0.094$ & This work \\ \hline
         $f_{gas, r500}$ + SN ($K^{\text{CMB}}\gamma^{\text{The300}}h^{\text{Planck}}$) & $-0.061 \pm 0.104$ & This work \\ \hline
         $f_{gas, r500}$ + SN ($K^{\text{Clash}}\gamma^{\text{The300}}h^{\text{Planck}}$) & $-0.047 \pm 0.116$ & This work \\ \hline
         $f_{gas, r500}$ + SN ($K^{\text{Clash}}\gamma^{\text{FABLE}}h^{\text{Planck}}$) & $-0.035 \pm 0.118$ & This work \\ \hline
         $f_{gas, r500}$ + SN ($K^{\text{CCCP}}\gamma^{\text{The300}}h^{\text{Planck}}$) & $-0.040 \pm 0.094$ & This work \\ \hline
         $f_{gas, r2500}$ + BAO ($h^{\text{SH0ES}}$) & $0.043 \pm 0.133$ & This work \\ \hline
         $f_{gas, r2500}$ + BAO ($h^{\text{Planck}}$)) & $0.081 \pm 0.137$  & This work \\ \hline
         $f_{gas, r2500}$ + SN ($h^{\text{SH0ES}}$) & $-0.019 \pm 0.136$ & This work \\ \hline
         $f_{gas, r2500}$ + SN ($h^{\text{Planck}}$) & $-0.059 \pm 0.133$  & This work \\ \hline

    \end{tabular}
    \caption{Values obtained from other articles that used a similar methodology and our results. The parametrization from all these works is the model where there is a constant deviation, i.e., $\epsilon(z) = \epsilon_0$. }
    \label{tabResultsComparacao}
\end{table}

\section{Conclusions}

In recent years, with the substantial volume of observations being conducted, tensions within the $\Lambda$CDM model have accumulated, highlighting the need for standard model modifications (or extensions) to reconcile the theory with observational measurements. One approach to addressing a possible departure from the standard model is to analyze the behavior of the dark matter energy density evolution. In this paper, we explore a potential interaction in the dark sector of the universe by allowing deviation in the standard dark matter density evolution through the phenomenological parametrizations of the variable interaction parameter $\epsilon(z)$ ($\rho_{{dm}}\propto(1+z)^{3 + \epsilon(z)}$, if $\epsilon(z)=0$, the standard model is recovered). We phenomenologically propose three $\epsilon(z)$-functions and tested them through the galaxy cluster gas mass fraction measurements, BAO, CC, and SNe Ia observations.

To quantify the performance of these functions concerning the datasets, we employed a Bayesian model selection with the constant $\epsilon(z)$ model as the reference model. All the parameterizations investigated in this study are listed in Table \ref{TabParametrizations}.  The results obtained for the parameterizations, consistent with the standard model, hold significant implications for our understanding of interaction in the dark sector. However, due to the uncertainties in the data, our results point to no conclusive position can be established. In the event of a deviation, the Bayes factors indicate either inconclusive evidence or a slight preference for the reference model. This analysis, which underscores the importance of our parameterization approach, is summarized in Tables \ref {TabR500BAO}-\ref {TabR2500SN}. It is important to comment that two different values for the sound horizon distance were used in BAO data (those obtained from the H0LiCOW+SN+BAO+SH0ES program and  Planck results \cite{aghanim2020planck}.
Moreover, two values for the supernova absolute magnitude ($M_B$) were also considered to obtain luminosity distance data (those from the SH0ES team and Planck results). In both analyses, data from H(z) cosmic chronometers were used to estimate values of $D_A(z)$ or $D_L(z)$, via Eq. (\ref{EqD_C}), thereby making the analysis more robust. However, our results were statistically unable to capture the cosmic tension in the Hubble parameter. 

It is worth emphasizing that more restrictive limits can be obtained on the interaction parameter employing perturbative equations and LSS and CMB data in the context of constant interaction. Such an analysis used the set of data SNIa+BAO+H(z)+LSS+CMB and the theoretical motivation for the interaction, which is based on the fundamental ground \cite{shapiro2002scaling, shapiro2000scaling,GmezValent2018, sola2023running}. However, the development presented here offers the possibility of conducting tests through an original methodology involving different observables and employing a model-independent approach.
It underscores the need for further research and more precise data on galaxy clusters. We hope that future projects, e.g., the eROSITA telescope, which promises to significantly increase the quantity and quality of data, will enable a more concise analysis of the topic developed here. Moreover, CMB and LSS data will be considered using the perturbative approach.

\vspace{0.5cm}

\section*{Acknowledgements}
The authors thank the contribution provided by the insightful comments and suggestions of the anonymous referee, which significantly improved the quality of the manuscript. Z. C. Santana Junior and M. O. Costa acknowledge the Coordenação de Aperfeiçoamento de Pessoal de Nível Superior (CAPES) for financial support. R. F. L. Holanda and R. Silva acknowledge the financial support from the Conselho Nacional de Desenvolvimento Científico e Tecnológico (CNPq).

\newpage

\bibliographystyle{ieeetr}
\bibliography{sample.bib}

\end{document}